\documentclass{article}
\usepackage{amssymb}
\usepackage{amsmath}

\input{tcilatex}

\begin{document}

\begin{center}
\textbf{STABILITY\ OF\ CIRCULAR\ ORBITS\ IN GENERAL RELATIVITY:\ A\ PHASE
SPACE ANALYSIS}

A. Palit$^{1,3,a}$, A. Panchenko$^{4,b}$, N.G. Migranov$^{4,c}$, A. Bhadra$%
^{2,3,d}$ and K.K. Nandi$^{1,3,4,e}$

$^{1}$Department of Mathematics

$^{2}$High Energy and Cosmic Ray Research Center

$^{3}$University of North Bengal, Siliguri 734 013, India.

$^{4}$Joint Research Laboratory, Bashkir State Pedagogical University, Ufa
450000, Russia.

$^{a}$Email: mail2apalit@gmail.com

$^{b}$Email: alex.souljaa@gmail.com

$^{c}$Email: ufangm@yahoo.co.uk

$^{d}$Email:\ aru\_bhadra@yahoo.com

$^{e}$Email: kamalnandi1952@yahoo.co.in

\bigskip

\textbf{Abstract}
\end{center}

Phase space method provides a novel way for deducing qualitative features of
nonlinear differential equations without actually solving them. The method
is applied here for analyzing stability of circular orbits of test particles
in various physically interesting environments. The approach is shown to
work in a revealing way in Schwarzschild spacetime. All relevant conclusions
about circular orbits in the Schwarzschild-de Sitter spacetime are shown to
be remarkably encoded in a \textit{single} parameter. The analysis in the
rotating Kerr black hole readily exposes information as to how stability
depends on the ratio of source rotation to particle angular momentum. As a
wider application, it is exemplified how the analysis reveals useful
information when applied to motion in a refractive medium, for instance,
that of optical black holes.

\begin{center}
\ \ \ \ \ \ \ \ \ \ \ \ \ \ \ \ \ \ \ \ \ \ \ \ \ \ \ \ \ \ \ \ \ \ \ \ \ \
\ \ \ \ \ \ \ \ \ \ \ \ \ \ \ \ \ \ \ \ \ \ \ \ \ \ \ \ \ \ \ \ \ \ \ \ \ \
\ \ \ \ \ \ \ \ \ \ \ \ \ \ \ \ \ \ \ \ \ \ \ \ \ \ \ \ \ \ \ \ \ \ \ \ \ \
\ \ \ \ \ \ \ \ \ \ \ \ \ \ \ \ \ \ \ \ \ \ \ \ \ \ \ \ \ 

\textbf{1. Introduction}
\end{center}

It is well known that in Einstein's theory of General Relativity (GR) motion
in a gravitational field is described by a system of dynamical equations.
The nonlinear ordinary differential equation for the path is obtained by
eliminating the affine parameter from that system of equations. An important
class of solutions of the path equation is formed by circular trajectories.
The issue of their stability is particularly important in confinement
problems and/or in accretion phenomenon in astrophysics. However,
conventional analysis of stability of orbits essentially deals with \textit{%
dynamical} equations involving the affine parameter and a potential
function. On the other hand, potential functions may not always be
immediately evident, for instance, in a simulated environment like moving
refractive dielectrics whereas path equations in them could follow directly
from Fermat's principle or Hamilton-Jacobi equation. Therefore, a natural
query is to ask if information about stability of circular orbits can be
obtained from the \textit{geometrical} path equation alone. Our aim here is
to rigorously demonstrate that it is indeed possible via phase space
analysis of autonomous systems corresponding to various physically
interesting environments. To our knowledge, such a useful application in
gravitational physics seems yet unavailable in the literature.

For our analysis, we shall not require any other information beyond the path
equation. The conserved angular momentum appearing in it will allow us to
connect the phase space results with actual kinematics. An equilibrium state
corresponds to a constant solution of a differential equation describing a
mechanical system and conversely. Constant solution means that velocity $%
\overset{.}{x}$ and acceleration $\overset{..}{x}$ be simultaneously zero.
The concept of stability of an equilibrium state is borrowed from the
familiar example of motion of a pendulum about the equilibrium point $x=0$
and $\overset{.}{x}=0$ where $x$ is angle with the vertical. The motion is
stable because a small displacement from the equilibrium position will lead
to only small oscillations of the bob about that position in a vertical
plane. Different closed paths on the phase space about a stable equilibrium
point correspond to real oscillations with different periods. If the paths
around an equilibrium point is such that a small displacement from the
equilibrium state takes the system far away from that point, it is called an
unstable equilibrium point. For instance, the equilibrium point $x=\pi $ and 
$\overset{.}{x}=0$ is an unstable saddle in the pendulum motion. There is no
closed path around such equilibrium point although some paths may pass
through it depending on the values of the parameter. Open paths not passing
through an equilibrium point represent whirling motion of the pendulum [1].

In this paper, we shall portray path equations for different solutions of GR
as plane autonomous dynamical systems and study them through phase space
and/or Hamiltonian analysis. Of particular interest is the treatment in
cosmological scenario and in the Kerr metric. As a further example, the
method is applied to light propagation in a medium relevant to optical black
holes and interesting information obtained.

The article is intended for theoretical physicists in general and practising
relativists in particular. The contents are organized as follows. In \textbf{%
Sec.2}, we briefly describe the path equation needed for our purpose. 
\textbf{Sec.3} treats the path equation as an autonomous system without the
cosmological constant $\lambda =0$. \textbf{Sec.4} treats the autonomous
system as a Hamiltonian system. \textbf{Sec.5} deals with the case $\lambda
\neq 0$. In \textbf{Sec.6}, we discuss circular motion in Kerr spacetime. 
\textbf{Sec.7} shows an example applying the method to light motion in a
refractive medium relevant to optical black holes. \textbf{Sec.8} summarizes
the obtained results. We take $G=c_{0}=1$, unless specifically restored.

\begin{center}
\textbf{2. Path equation}
\end{center}

A spherically symmetric static solution of the Einstein field equations for
a mass $M$ is given by the Schwarzschild-de Sitter solution (SdS) in
standard coordinates ($x^{\alpha }$)$\equiv $($t,r,\theta ,\phi $):%
\begin{eqnarray}
d\tau ^{2} &=&B(r)dt^{2}-B^{-1}(r)dr^{2}-r^{2}d\theta ^{2}-r^{2}\sin
^{2}\theta d\phi ^{2}, \\
B(r) &=&1-\frac{2M}{r}-\frac{\lambda r^{2}}{3}  \notag
\end{eqnarray}%
where $\lambda >0$ is the cosmological constant. The cosmological constant $%
\lambda \sim 10^{-55}cm^{-2}>0$ is responsible \ for dark energy that
explains the currently observed accelerated cosmic expansion [2]. The case $%
\lambda <0$ (anti-de Sitter) is observationally ruled out and we shall not
deal with this case here. The case $M\neq 0$, $\lambda =0$ corresponds to
the pure Schwarzschild solution and $r_{H}=2M$ is the horizon radius so that
the metric is valid for $r>r_{H}$. $\ $When $M=0$, $\lambda \neq 0$ we have
pure de Sitter solution which can be reexpressed in such a manner that it
represents an expanding space devoid of matter.

When $M\neq 0$, $\lambda \neq 0$, there occur two horizons provided that $%
0<\lambda <\lambda _{crit}=\frac{1}{9M^{2}}$. The black hole horizon appears
at $r_{h}=\frac{1}{\sqrt{\lambda M^{2}}}\cos \frac{\pi +\xi }{3}$ and the
cosmological horizon at $r_{c}=\frac{1}{\sqrt{\lambda M^{2}}}\cos \frac{\pi
-\xi }{3}$ where $\xi =\cos ^{-1}(\sqrt{3}\lambda M^{2})$. The spacetime is
dynamic for $r<r_{h}$ and for $r>r_{c}$. At the critical value $\lambda
=\lambda _{crit}$, the two horizons coincide at $r_{ph}=3M$. The static
radius $r_{st}$ is defined as a hypersurface where the attraction due to $M$
\ balances the cosmic repulsion due to $\lambda $ and is given by [3]%
\begin{equation}
r_{st}=\left( \frac{3M}{\lambda }\right) ^{\frac{1}{3}}.
\end{equation}%
All circular orbits are bounded from below by $r=r_{ph}$ and from above by
the static radius $r=r_{st}$.

Following conventional stability analyses of circular orbits involving a
potential function $V(r)$, we obtain the expression%
\begin{equation}
\frac{d^{2}V}{dr^{2}}\mid _{r=R}=-\left( \frac{2M}{R^{3}}\right) \left( 
\frac{\frac{4}{3}\lambda R^{4}-5\lambda MR^{3}+R-6M}{R-3M}\right) .
\end{equation}%
In the case $\lambda =0$, it follows from Eq.(3) that stable circular orbits
may exist only at radii $R>6M$. At $R=6M$, $\frac{d^{2}V}{dr^{2}}\mid
_{r=R}=0$ which indicates that it is a point of inflection. When $%
R\rightarrow +\infty $ and $3M+$, we have $\frac{d^{2}V}{dr^{2}}\rightarrow
0-$ and $+\infty $ respectively. The first limit indicates stability in the
asymptotic region while the divergent second limit indicates instability.

Defining $u=\frac{1}{r}$, we have the path equation on the equatorial plane $%
\theta =\pi /2$ for a particle as%
\begin{equation}
\overset{..}{x}=a+bx+cx^{2}+dx^{-3}
\end{equation}%
where an overdot denotes differentiation with respect to $\phi $ and%
\begin{eqnarray}
x &=&u=\frac{1}{r} \\
a &=&\frac{M}{h^{2}} \\
b &=&-1 \\
c &=&3M \\
d &=&-\frac{\lambda }{3h^{2}} \\
h &=&r^{2}\frac{d\phi }{d\tau }=const.
\end{eqnarray}%
The quantity $h$ is the conserved angular momentum. \textit{Eq.(4) with the
defined coefficients is all we need.}

We shall study it in the phase plane ($x=u,y\equiv \frac{du}{d\phi }$)
posing it as an autonomous system as follows 
\begin{eqnarray}
\overset{.}{x} &=&y \\
\overset{.}{y} &=&a+bx+cx^{2}+dx^{-3}.
\end{eqnarray}%
Equilibrium points of the system are given by 
\begin{equation}
\overset{.}{x}=0,\overset{.}{y}=0.
\end{equation}%
The first equation gives circular orbits $r=R=const$. and the second
equation gives the angular momentum $h$ along that orbit 
\begin{equation}
h^{2}=\frac{MR(1-\frac{\lambda }{3M}R^{3})}{1-\frac{3M}{R}}.
\end{equation}%
It follows that $h^{2}=0$ at $R=r_{st}$ and $h^{2}=\infty $ at $R=3M$. Let
us now consider the case $\lambda =0$ corresponding to the Schwarzschild
spacetime.

\begin{center}
\textbf{3. Case I: }$\lambda =0$
\end{center}

The autonomous system%
\begin{eqnarray}
\overset{.}{x} &=&y \\
\overset{.}{y} &=&a+bx+cx^{2}
\end{eqnarray}%
gives two equilibrium points ($x,y$) at%
\begin{equation*}
P:\left( \frac{-b+\sqrt{b^{2}-4ac}}{2c},0\right) ;Q:\left( \frac{-b-\sqrt{%
b^{2}-4ac}}{2c},0\right) .
\end{equation*}%
In order to have these points located on the real phase plane, we have to
assume $b^{2}-4ac\geq 0$ or rephrasing, $\alpha ^{2}\equiv 1-\frac{12M^{2}}{%
h^{2}}\geq 0$. We have introduced the shorthand $\alpha $ for notational
convenience. Each choice of $\alpha $ gives a corresponding equilibrium
point or a value of the radius $r$. Let us first consider the degenerate
case $\alpha =0$ and study the stability of corresponding radius.

\textit{Case (a): }$k^{2}\equiv b^{2}-4ac=0\Rightarrow a=\frac{b^{2}}{4c}%
\neq 0.\ $

The equilibrium point on the phase plane occurs only at ($-\frac{b}{2c},0$).
Eliminating the parameter $a$, the autonomous system (15), (16) can be
reduced to the following set of equations%
\begin{eqnarray}
\overset{.}{x} &=&y \\
\overset{.}{y} &=&\frac{1}{4c}\left( b+2cx\right) ^{2}.
\end{eqnarray}%
The differential phase path is given by%
\begin{equation}
\frac{dy}{dx}=\frac{(b+2cx)^{2}}{4cy}
\end{equation}%
which integrates to 
\begin{equation}
y^{2}=\frac{1}{12c^{2}}(b+2cx)^{3}+A
\end{equation}%
where $A$ is an arbitrary constant of integration.

We see that the parameter $b$ has the effect of only translation in the
variable $x$ while $c$ introduces magnification in both $x$ and $y$. Thus
the GR correction term $c$ can be regarded as the dominating parameter among 
$a,b$ and $c$. By the translation%
\begin{eqnarray}
y^{\prime } &=&y \\
x^{\prime } &=&x+\frac{b}{2c}
\end{eqnarray}%
the autonomous system (15), (16) further reduces to%
\begin{eqnarray}
\overset{.}{x}^{\prime } &=&y^{\prime } \\
\overset{.}{y}^{\prime } &=&cx^{\prime 2}
\end{eqnarray}%
which gives a one-parameter family of phase paths%
\begin{equation}
3y^{\prime 2}=2cx^{\prime 3}+C
\end{equation}%
on the phase plane ($x^{\prime },y^{\prime }$) where $C$ is an arbitrary
parameter. The phase paths for $M=1$ (or $c=3$) and different values of $C$
are given in Fig.1.

The equilibrium point has now been shifted to the origin ($0,0$) which gives%
\begin{eqnarray}
x^{\prime } &=&0\Rightarrow r=6M\Rightarrow hu\equiv r\frac{d\phi }{d\tau }%
=const. \\
y^{\prime } &=&0\Rightarrow -h\frac{du}{d\phi }\equiv \frac{dr}{d\tau }=0.
\end{eqnarray}%
From the above, we immediately learn the following:\ The equilibrium point
corresponds, in the physical ($r,\phi $) plane, to a circular orbit of
radius $6M$ with the test particle having a constant cross radial velocity $%
hu$ and a zero radial velocity $-h\frac{du}{d\phi }$. From the overall
pattern of the phase paths given in Fig.1, we see that a small displacement
from the equilibrium state can take the system on a phase path which leads
it far away from the equilibrium state. The dynamical condition for this to
happen is given by%
\begin{equation}
\alpha ^{2}=0\Rightarrow h^{2}=12M^{2}.
\end{equation}%
Although the phase path Eq.(25) is independent of $a$, it applies only to
massive test particles because the value of $h^{2}$ becomes infinity for
light ($d\tau =0$). In this case, the condition (28) becomes obviously
inapplicable. We shall treat this case separately in Sec.4.

Let us analyze in a little more detail the paths in different quadrants in
Fig.1. A typical initial state ($x_{0}^{\prime },y_{0}^{\prime }$) on the
phase plane is as follows%
\begin{eqnarray}
x_{0}^{\prime } &=&x_{0}+\frac{b}{2c}=x_{0}-\frac{1}{6M}=\delta \\
y_{0\pm }^{\prime } &=&y_{0\pm }=\pm \sqrt{\frac{6x_{0}^{\prime 3}+C}{3}}%
=\pm \sqrt{\frac{6\delta ^{3}+C}{3}}
\end{eqnarray}%
where $C>-6\delta ^{3}$. These equations will allow us to closely examine
phase paths in the neighborhood of the equilibrium point. We see from
Eq.(29) that the GR allowed open interval $x_{0}^{-1}=r_{0}\in (2M,+\infty )$
is mapped onto a finite open interval $x_{0}^{\prime }\in (-\frac{1}{6M},%
\frac{1}{3M})$ around the equilibrium point ($0,0$). This interval can be
subdivided into two parts for $\delta $ or $x_{0}^{\prime }$:

One part is $\delta \in \lbrack 0,\frac{1}{3M})$ corresponding to $6M\geq
r_{0}>2M$. This interval refers to points ($x_{0}^{\prime },y_{0+}^{\prime }$%
) on the paths in first quadrant and to points ($x_{0}^{\prime
},y_{0-}^{\prime }$) on the paths in the fourth quadrant. To proceed
further, let us translate Eqs.(29),(30) to the physical ($r,\phi $) plane
choosing $C=0$: 
\begin{eqnarray}
r_{0} &=&\frac{6M}{6M\delta +1} \\
\frac{du}{d\phi } &\mid &_{0\pm }=\pm \sqrt{2\delta ^{3}}\Rightarrow \frac{dr%
}{d\tau }\mid _{0\pm }\equiv -h\left( \frac{du}{d\phi }\right) _{0\pm }=\mp h%
\sqrt{2\delta ^{3}}.
\end{eqnarray}%
(Note that the cross radial component of velocity can be expressed as$\ r%
\frac{d\phi }{d\tau }=hx^{\prime }$ and the radial component as $\frac{dr}{%
d\tau }=-hy^{\prime }$.) We find the following distinct possibilities for
paths (25) passing through the equilibrium point ($0,0$): (i) As $\delta $
increases from $0$ to $\frac{1}{3M}$, we see from Eqs.(29), (30) that both $%
x_{0}^{\prime }$ and $y_{0+}^{\prime }$ increase from the equilibrium point (%
$0,0$) which implies that the phase point in the first quadrant moves
outward (to the right), as represented by $OB$ in Fig.1. Correspondingly,
from Eq.(32), we see that $u$ increases with $\phi $ which indicates that
the radius $r_{0}$ undergoes a decrease in time $\tau $ from $6M$ to $2M$
(as reflected in $\frac{dr}{d\tau }\mid _{+}=-h\sqrt{2\delta ^{3}}<0$). (ii)
As $\delta $ decreases from $\frac{1}{3M}$ to $0$, we see that both $%
x_{0}^{\prime }$ and $y_{0-}^{\prime }$ decrease to the equilibrium point ($%
0,0$) which implies that the phase point in the fourth quadrant moves inward
(to the left), as represented by $AO$ in Fig.1$.$ Correspondingly, $u$
decreases with $\phi $ and the radius $r_{0}$ undergoes an increase in time $%
\tau $ from $2M$ to $6M$ (as reflected in $\frac{dr}{d\tau }\mid _{\_}=+h%
\sqrt{2\delta ^{3}}>0$). For $C\neq 0$, paths will not pass through ($0,0$)
but parts of $CDE$ lying in the first and fourth quadrant can be interpreted
similarly.

The other part is $\delta \in (-\frac{1}{6M},0]$, for which $\infty
>r_{0}\geq 6M$. This represents points only on the second and third quadrant
(where $x_{0}^{\prime }=\delta <0$). In this case, to avoid imaginary
quantity in Eq.(32), we must choose $C\neq 0$, that is, we have to deal with
the full set of Eqs. (29), (30). For different values of $C\neq 0$ in Fig.1,
we see that the phase paths like $CDE$ are not closed around the equilibrium
point. (These paths are analogous to whirling motion of pendulum). It is
clear that most of the paths in the interval $-\frac{1}{6M}\leq
x_{0}^{\prime }<\frac{1}{3M}$, when slightly displaced from the equilibrium
point ($0,0$), neither converge to it nor form a center about it. In fact,
the phase paths resemble exactly those around a \textit{cusp }[1]\textit{. }%
This leads us to conclude that the equilibrium point ($0,0$) corresponding
to radius $r=6M$ is neither stable or nor unstable because of the
dynamically degenerate condition $\alpha =0$. This conclusion will be
further supported in Sec.4.

\bigskip \textit{Case (b)}: $k^{2}\equiv b^{2}-4ac>0$

There are now two distinct equilibrium points occurring at ($\frac{-b+k}{2c}%
,0$) and ($\frac{-b-k}{2c},0$). They combine into a single representative
point ($\frac{-b+\alpha }{2c},0$) where $\alpha =+k$ or $-k$. Under the
translation%
\begin{eqnarray}
y^{\prime } &=&y \\
x^{\prime } &=&x-\frac{\alpha -b}{2c}
\end{eqnarray}%
the autonomous system (15), (16) reduces, after a little algebra, to%
\begin{eqnarray}
\overset{.}{x}^{\prime } &=&y^{\prime } \\
\overset{.}{y}^{\prime } &=&\alpha x^{\prime }+cx^{\prime 2}.
\end{eqnarray}%
The equilibrium points in the new ($x^{\prime },y^{\prime }$) phase plane
are $P_{1}:$($0,0$) and $Q_{1}:$($-\frac{\alpha }{c},0$). \ The linearized
system of equations near $P_{1}:$($0,0$) is%
\begin{eqnarray}
\overset{.}{x}^{\prime } &=&y^{\prime } \\
\overset{.}{y}^{\prime } &=&\alpha x^{\prime }.
\end{eqnarray}%
Comparing it with the general linear system given by%
\begin{eqnarray}
\overset{.}{x}^{\prime } &=&a_{1}x^{\prime }+b_{1}y^{\prime } \\
\overset{.}{y}^{\prime } &=&c_{1}x^{\prime }+d_{1}y^{\prime }
\end{eqnarray}%
we find 
\begin{equation}
a_{1}=0,b_{1}=1,c_{1}=\alpha
,d_{1}=0,p=a_{1}+d_{1}=0,q=a_{1}d_{1}-b_{1}c_{1}=-\alpha
\end{equation}%
so that the discriminant is 
\begin{equation}
\Delta \equiv p^{2}-4q=4\alpha .
\end{equation}%
Hence the equilibrium point ($0,0$) will be a center (stable equilibrium) if 
$\alpha <0$ and a saddle point (unstable equilibrium) if $\alpha >0$. Such
an abrupt change in the behavior of the system occurs through $\alpha =0$.
Therefore, $\alpha =0$ can be called a \textit{bifurcation }point.

The above conclusions are supported by the phase paths following from
Eqs.(35) and (36), namely,%
\begin{equation}
y^{\prime 2}=\alpha x^{\prime 2}+\frac{2c}{3}x^{\prime 3}+D
\end{equation}%
where $D$ is an arbitrary constant of integration. In the close vicinity of (%
$0,0$) such that $x^{\prime }\sim 0$ and $x^{\prime 3}$ can be neglected, we
get%
\begin{equation}
y^{\prime 2}-\alpha x^{\prime 2}=D
\end{equation}%
which represents a family of concentric ellipses for $\alpha =-k<0$ (Fig.2,
center, stable) and a family of hyperbolas with asymptotes $y^{\prime }=\pm 
\sqrt{\alpha }x^{\prime }$ for $\alpha =+k>0$ (Fig.3, saddle point,
unstable). Note that $\alpha \sim 1$, since, for physically realistic
particle orbits, $\frac{M^{2}}{h^{2}}\ll 1$. Thus, as a sample, we have
taken $\alpha =\pm k=\pm 0.9,$ and $c=3$ (which means we are taking units in
which $M=1$) and different values of the parameter $D$ in Figs.2,3.

Let us see what conclusions we can draw in the physical ($r,\phi $) plane.
The point ($x^{\prime },y^{\prime }$)$\equiv $($0,0$) shows that the
equilibrium radii $r$ depend on the value of $h$, and hence of $\alpha $.
These radii follow from Eq.(34) 
\begin{equation}
x^{\prime }=x-\frac{\alpha -b}{2c}=0\Rightarrow r=\frac{6M}{1+\alpha }.
\end{equation}%
This means $\infty >r>6M$ if $-1<\alpha <0$ $\ $and $6M>r>2M$ if $0<\alpha
<2 $. Thus, from what we have learnt from Eq.(44), we find that circular
orbits with $r>6M$ are stable ($\alpha $ has negative sign) while those with 
$r<6M$ are unstable ($\alpha $ has positive sign). Eq.(33) gives 
\begin{equation}
y^{\prime }=y=0\Rightarrow \frac{du}{d\phi }=0\Rightarrow \frac{dr}{d\tau }%
=-h\frac{du}{d\phi }=0
\end{equation}%
while $r\frac{d\phi }{d\tau }\equiv hu>0$. That is, the radius $r$ is
independent of time while the cross radial velocity $r\frac{d\phi }{d\tau }$
is nonzero. These are exactly what are to be expected of circular motions.

The other equilibrium point ($x^{\prime },y^{\prime }$)$\equiv $($-\frac{%
\alpha }{c},0$) corresponds to 
\begin{equation}
-\frac{\alpha }{c}=x-\frac{\alpha -b}{2c}\Rightarrow r=\frac{6M}{1-\alpha }.
\end{equation}%
In the close vicinity of this equilibrium point, we may define%
\begin{equation}
x^{\prime \prime }=x^{\prime }+\frac{\alpha }{c}.
\end{equation}%
When $x^{\prime \prime }\sim 0$, neglecting $x^{\prime \prime 2}$, we have
from Eqs. (35), (36), the linearized system%
\begin{eqnarray}
\overset{.}{x}^{\prime \prime } &=&y^{\prime \prime } \\
\overset{.}{y}^{\prime \prime } &=&-\alpha x^{\prime \prime }.
\end{eqnarray}%
Arguing in the same manner as with Eqs.(37) and (38), we see that we have
here a \textit{reverse }situation, viz., the point $Q_{1}:$($-\frac{\alpha }{%
c},0$) is a saddle for $\alpha =-k$ $<0$ and a center for $\alpha =+k>0$. We
shall show below that this is indeed the case. For this, we shall
investigate stability by posing the autonomous system as a geometrical
Hamiltonian system. The latter technique is said to be more reliable than
the linearization technique [1].

\begin{center}
\textbf{4. Hamiltonian system}
\end{center}

Dropping primes in the autonomous system, Eqs.(35), (36), the Hamiltonian
system can be defined as%
\begin{eqnarray}
\frac{\partial H}{\partial x} &=&-Y(x,y)=-(\alpha x+cx^{2}) \\
\frac{\partial H}{\partial y} &=&X(x,y)=y
\end{eqnarray}%
The necessary and sufficient condition for the system (51), (52) to be a
Hamiltonian system, namely, $\frac{\partial X}{\partial x}+\frac{\partial Y}{%
\partial y}=0$, is fulfilled for all $x$ and $y$. [Such fulfillment is a
special feature of the GR path Eq.(4)]. Moreover, $\frac{dH}{d\phi }=0$ and
therefore $H(x,y)=$const. independent of $\phi $. From the Eqs.(51) and
(52), we get%
\begin{eqnarray}
H(x,y) &=&-\frac{\alpha }{2}x^{2}-\frac{c}{3}x^{3}+u(y) \\
H(x,y) &=&\frac{1}{2}y^{2}+v(x)
\end{eqnarray}%
where $u(y)$ and $v(x)$ are arbitrary functions subject to the consistency
of Eqs.(53) and (54). These two equations will match only if%
\begin{eqnarray}
u(y) &=&\frac{1}{2}y^{2}-C \\
v(x) &=&-\frac{\alpha }{2}x^{2}-\frac{c}{3}x^{3}-E
\end{eqnarray}%
where $E$ is an arbitrary constant. The Hamiltonian paths are given by%
\begin{equation}
H(x,y)=-\frac{\alpha }{2}x^{2}-\frac{c}{3}x^{3}+\frac{1}{2}y^{2}-E
\end{equation}%
where $E$ is a parameter. It follows that

\begin{eqnarray}
\frac{\partial ^{2}H}{\partial x^{2}} &=&-(\alpha +2cx) \\
\frac{\partial ^{2}H}{\partial y^{2}} &=&1 \\
\frac{\partial ^{2}H}{\partial x\partial y} &=&0.
\end{eqnarray}%
As before, the equilibrium points occur when $X=0$ and $Y=0$ which give the
points $P_{1}:(0,0)$ and $Q_{1}:(-\frac{\alpha }{c},0).$ Thus the quantity%
\begin{equation}
q_{0}\equiv \frac{\partial ^{2}H}{\partial x^{2}}\frac{\partial ^{2}H}{%
\partial y^{2}}-\left( \frac{\partial ^{2}H}{\partial x\partial y}\right)
^{2}
\end{equation}%
has the following values%
\begin{eqnarray}
q_{0} &\mid &_{P_{1}}=-\alpha \\
q_{0} &\mid &_{Q_{1}}=\alpha .
\end{eqnarray}%
When $-1<\alpha <0$, the equilibrium point $P_{1}$ is a stable center since $%
q_{0}>0$, but $Q_{1}$ is an unstable saddle point. For $\alpha >0$, the
conclusions are reversed. These confirm the results of Sec.3. The value $%
\alpha =0$ is a bifurcation point as it represents a transition of the
system from a stable center to an unstable saddle and conversely.

We shall now see what result do we get applying the present method to light
trajectories for which $a=0$. From the original set of Eqs. (15), (16), we
see that they lead to the same Hamiltonian set of Eqs.(51), (52) with the
difference that $\alpha $ is now to be replaced by $b$. The equilibrium
points then are $P_{2}:$($0,0$) and $Q_{2}:$($-\frac{b}{c},0$). The point $%
P_{2}$ implies%
\begin{equation}
x=0\Rightarrow r\rightarrow \infty
\end{equation}%
and the value of $q_{0}\mid _{P_{2}}=-b=1>0.$ This implies that $P_{2}$ is a
center. From this, we learn that light trajectories (straight lines) in
asymptotically flat space ($r\rightarrow \infty $) are stable. This is an
expected result. The other equilibrium point $Q_{2}$ implies 
\begin{equation}
x=-\frac{b}{c}\Rightarrow r=3M
\end{equation}%
at which $q_{0}\mid _{Q_{2}}=b=-1<0$ showing that $Q_{2}$ represents a
saddle point. In other words, light orbit at $r=3M$ is unstable. This shows
that the instability of circular orbits of light at $R=3M$ depends only on
the sign of $b$ and is \textit{independent} of the sign of $\alpha $, unlike
in the case of material orbits.

\begin{center}
\textbf{5. Case II: }$\lambda \neq 0$
\end{center}

We have to consider the full autonomous system (11), (12) and as usual, the
equilibrium points are given by $\overset{.}{x}=0,$ $\overset{.}{y}=0$. The
latter gives the equation%
\begin{equation}
g(x)\equiv cx^{5}+bx^{4}+ax^{3}+d=0,x\neq 0.
\end{equation}%
Since $x$ has to be non-negative, we have to look only for positive roots of 
$g(x)=0$. Once the known signs of coefficients are plugged into $g(x)=0$, we
may apply Descartes' rule of signs to see that $g(x)=0$ can have either one
or three positive real roots, the rest are either negative or imaginary. The
auxiliary equation $g^{\prime }(x)=\frac{dg}{dx}=0$ has two zero roots and
two nonzero roots $\mu _{1}$, $\mu _{2}$ given by%
\begin{equation}
\mu _{1}=\frac{-2b-\sqrt{4b^{2}-15ac}}{5c},\mu _{2}=\frac{-2b+\sqrt{%
4b^{2}-15ac}}{5c}.
\end{equation}%
The reality of the roots of $g^{\prime }(x)=0$ demands that $\gamma $ $%
^{2}\equiv 4b^{2}-15ac\geq 0$. Since $b=-1$ and $c>0$, we see that $\mu
_{1}\leq \mu _{2}$. Let us denote a representative positive root of $g(x)$
by $\eta \neq 0$, that is, $g(\eta )=0$. The representative equilibrium
point is then ($x,y$) = ($\eta ,0$). Then we employ the usual operations on
Eqs.(11) and (12), viz., a\ translation $x^{\prime }=x-\eta $, $y^{\prime
}=y $, followed by linearization in the neighborhood of ($x^{\prime
},y^{\prime } $) = ($0,0$). The final result is%
\begin{eqnarray}
\overset{.}{x}^{\prime } &=&y^{\prime } \\
\overset{.}{y}^{\prime } &=&\left( \frac{5c\eta ^{2}+4b\eta +3a}{\eta }%
\right) x^{\prime }.
\end{eqnarray}%
Using Eqs.(39)-(42), we get 
\begin{equation}
q=-\left( \frac{5c\eta ^{2}+4b\eta +3a}{\eta }\right) ,p=0,\Delta =4\left( 
\frac{5c\eta ^{2}+4b\eta +3a}{\eta }\right) .
\end{equation}%
For a meaningful analysis, we must have $q\neq 0$ which means $g^{\prime
}(\eta )\neq 0$, that is, $\eta $ can not be a repeated root of $g(x)=0$.
Thus, we find that ($x^{\prime },y^{\prime }$) = ($0,0$) will be a saddle if 
$q<0$ and $\Delta >0$. This is possible if either $\eta <\mu _{1}$ or $\eta
>\mu _{2}$. The point ($x^{\prime },y^{\prime }$) = ($0,0$) will be a center
if $q>0$ and $\Delta <0$ which means $\mu _{1}<\eta <\mu _{2}$. The
linearization scheme is not applicable for $\eta =0$. The important point to
note here is that $\mu _{1},\mu _{2}$ do not depend on the cosmological
constant $\lambda $. Thus the constraint $\gamma ^{2}=4-\frac{45M^{2}}{h^{2}}%
\geq 0$ applies to orbits resulting from the effect of $M$ alone. Orbits
close to the static radius are not sensitive to this constraint.

With the above general picture in mind, let us numerically study the
behavior of approximate roots of $g(x)=0$ for some choices of $h$. For a
given $M$ the equilibrium points, hence the radii, vary depending on the
values of $h$, or $x_{eq}=x_{eq}(h,\lambda )$. Choosing units in which $M=1$%
, and with the values of coefficients given by $c=3,b=-1,a=h^{-2},d=-\frac{%
\lambda }{3h^{2}}$ the equation $g(x)=0$ can be rewritten as%
\begin{equation}
h^{2}(3x^{5}-x^{4})+x^{3}-\frac{\lambda }{3}=0;h\neq 0
\end{equation}%
We observe the following behavior. When $h^{2}\rightarrow 0$, we get only
one very small root that approximates to the static radius $%
x_{eq}=x_{st}=\left( \frac{\lambda }{3}\right) ^{\frac{1}{3}}\sim 10^{-18}$.
Other roots are imaginary. As we increase $h^{2}$ up to $\frac{45}{4}$, we
see that the picture remains almost the same, that is, we continue to obtain
a single radius of the order of $x_{st}$. When $h^{2}$ $>$ $\frac{45}{4}$ or 
$\gamma ^{2}>0$, we find that there occur three positive roots, one is of
the order of the same static radius, but the other two roots correspond to
orbits in the vicinity of $M$. These results confirm that the radii of
orbits close to or at the static radius are indeed insensitive to values of $%
\gamma $. Let us consider a specific value $h=8$ (say), then we have the
following equilibrium points:\ $P_{1}:(x=x_{st}$, $y=0)$, $P_{2}:(x=0.016$, $%
y=0)$ and $P_{3}:(x=0.316$, $y=0)$ while $\mu _{1}=0.012$, $\mu _{2}=0.254$.
\ According to the general discussion above, we expect that $P_{2}$ should
be a stable center as $\mu _{1}$ $<x<\mu _{2}$ while $P_{1}$and $P_{3}$
should be unstable saddles.

Let us confirm the results by the method of Hamiltonian system. Following
the same procedure as in Sec.4, we deduce that%
\begin{equation}
H(x,y)=\frac{1}{2}y^{2}-(ax+\frac{b}{2}x^{2}+\frac{c}{3}x^{3}-\frac{d}{2}%
x^{-2})-F.
\end{equation}%
where $F$ is an arbitrary parameter. The expression for $q_{0}$ is%
\begin{eqnarray}
q_{0} &=&-(b+2cx-3dx^{-4}) \\
&=&1-6x-\frac{\lambda }{h^{2}}x^{-4}.
\end{eqnarray}%
From this, we can conclude the following: At the lower bound, that is, at
the local photon radius $x_{ph}=1/3$, we find $q_{0}=-1$ since $h^{2}=\infty 
$. Therefore this particular orbit is unstable and the instability is
independent of $\lambda $. On the other hand, at the static radius, $%
q_{0}\mid _{x=x_{st}}>0$, implying that the photon orbit (again $%
h^{2}=\infty $) is stable at $x=x_{st}$. The stability of light orbits at
the static hypersurface is similar to that in the asymptotically flat region
discussed in Sec.4. At $x=\frac{1}{6}$, $q_{0}<0$, hence $R=6$ is also an
unstable radius. Furthermore, $q_{0}\mid _{P_{2}}>0$ and $q_{0}\mid
_{P_{3}}<0$ confirming earlier expectations. At the static radius $x=x_{st}$%
, we have $h^{2}=0$, and $\lambda x_{st}^{-4}\sim 10^{16}$ so that $%
q_{0}\mid _{P_{1}}=-\infty $. This shows that circular material orbit at the
static radius ($P_{1}$) is unstable.

What then is the upper bound $R_{ub}$ for stable circular material orbits?
This can be found by requiring that $q_{0}>0$ or 
\begin{equation}
6x+\frac{\lambda }{h^{2}}x^{-4}<1.
\end{equation}%
Putting the expression for $h^{2}$ from Eq.(14), and assuming that $%
R_{ub}\gg 6$, we find that 
\begin{equation}
R_{ub}=4^{-\frac{1}{3}}\left( \frac{\lambda }{3}\right) ^{-\frac{1}{3}}
\end{equation}%
which is slightly smaller than $r_{st}$. The radii at which orbits begin to
be stable can be obtained from $q_{0}=0$ which gives%
\begin{equation}
\lambda =h^{2}x^{4}(1-6x),x\neq 0
\end{equation}%
The maximum of $\lambda $ is located at $x=\frac{2}{15}$. Again using the
expression for $h^{2}$ from Eq.(14), we get 
\begin{equation}
\lambda _{\max }=\frac{4}{5625}\simeq 0.000711.
\end{equation}%
For $\lambda \sim 10^{-55}cm^{-2}$, the maximum Schwarzschild mass is $%
M_{\max }=\left( \frac{\lambda _{\max }}{\lambda }\right) ^{\frac{1}{2}}\sim
5.75\times 10^{20}M_{\odot }$. Stable material circular orbits can exist
only at $R>\frac{15}{2}$ corresponding to $\gamma >0$. This is confirmed by
the stability at $P_{2}$ ($R\simeq 10$) and instability at $P_{3}$($R\simeq
3.3$). The values $\lambda _{\max }$ and $R=\frac{15}{2}$ correspond to
another critical value $h^{2}=\frac{45}{4}$, as may be obtained from
Eq.(14). We have obtained it here from a totally different consideration,
namely, of roots of $g(x)=0$. For $h^{2}\geq \frac{45}{4}$, there exist
local stable equilibrium orbits $x_{eq}$ while no stable local $x_{eq}$
exist for $h^{2}<\frac{45}{4}$. We see that the restriction $h^{2}\geq \frac{%
45}{4}$ is \textit{weaker} than the previous $h^{2}\geq 12$ for the case $%
\lambda =0$.

It is remarkable that a \textit{single} parameter $q_{0}$ completely
reproduces all the results obtained by Stuchl\'{\i}k and Hled\'{\i}k [3],
including their numerical value, viz, $y_{c(ms)}=\frac{\lambda _{\max }}{3}%
=0.000237$.

The phase space method can also be applied in the pure de Sitter space which
corresponds to $\lambda >0$, $M=0$. There is now no balance of forces at any
radius, hence there is no static radius. The metric with $M=0$ immediately
fixes $a=c=0$ in $g(x)=0$. \ The equilibrium points then occur at $x^{4}=-%
\frac{\lambda }{3h^{2}}$. This implies that there are no real equilibrium
points and we conclude that circular orbits are not possible in this space.

\begin{center}
\textbf{6. Path equation in Kerr spacetime}
\end{center}

The phase space method can be profitably utilized in the study of motions in
a refractive medium as well. For instance, Evans and Rosenquist [4] showed
that the equation of optics in a refractive medium of index $n(%
\overrightarrow{r})$ can be effectively rephrased as a Newtonian
\textquotedblleft $\overrightarrow{f}=m\overrightarrow{a}\textquotedblright $
form of mechanics. This optical-mechanical analogy led via Fermat's
principle to a path equation for light in the form 
\begin{equation}
\frac{d^{2}\overrightarrow{r}}{dA^{2}}=\overrightarrow{\triangledown }\left( 
\frac{n^{2}}{2}\right)
\end{equation}%
where $A$ is a stepping parameter defined in Ref.[4] by $dA=n^{-2}dt$.
Optical analogues of mechanical quantities are marked by \textquotedblleft
..". The equation of motion (79) has been subsequently extended in Ref.[5]
to include also the motion of material particles. Note that the form of $n(%
\overrightarrow{r})$ can be arbitrarily preassigned depending on the nature
of the medium. The relevant quantities in this formalism are the optical
version of mechanical quantities. For instance, instead of the classical
angular momentum $h$, its optical analogue \textquotedblleft $h_{0}$", viz., 
$h_{0}=r^{2}\frac{d\phi }{dA}$ is conserved if $n=n(r)$. A specific form of $%
n=n(r)$ depicting a Schwarzschild gravitational \textquotedblleft medium"\
exactly yielded the path Eq.(4) for $\lambda =0$ [6]. (See also Ref.[7] for
another interesting derivation). It is clear that complicated forms of $n(r)$
corresponding to arbitrary spherical media would lead to path equations more
complicated than Eq.(4). In these cases, the present method might be
preferable to conventional methods.

An example is Kerr spacetime which represents a unique rotating black hole
solution for $\lambda =0$. Alsing [8] has extended the \textquotedblleft
medium" analogy to Kerr spacetime with rotation parameter $J$ (= angular
momentum per unit mass $M$ of the rotating source) and obtained, to first
order in $\frac{M}{r}$, the following path equations on the equatorial slice:%
\begin{eqnarray}
\frac{d^{2}u}{d\phi ^{2}}+u-3Mu^{2} &=&\frac{M}{L^{2}}\left[ \left( 1-\frac{%
v_{0}^{2}}{c_{0}^{2}}\right) \left( 1-8Mu\frac{J}{L}\right) -2\frac{J}{L}%
\left( \frac{v_{0}^{2}}{c_{0}^{2}}\right) \right] (part.) \\
&=&\frac{-2MJ}{L^{3}}(light)
\end{eqnarray}%
Here $v_{0}$ is the initial velocity of the particle at infinity and $L$ is
its conserved total \textquotedblleft angular momentum\textquotedblright\
per unit test mass given by%
\begin{equation}
L=\rho ^{2}\frac{d\phi }{dA}-\frac{2MJ}{\rho }
\end{equation}%
where $\rho =re^{-Mr}$. (In the asymptotic region, $n=1$, $A=t$ $\rho =r$ so
that, for $J=0$, we have $L=h$, the familiar mechanical angular momentum).
For a particle starting a radial fall from infinity, $L=0$. \ In this case,
with $n(r)\sim 1+\frac{2M}{\rho }$, one obtains to lowest order in $\frac{M}{%
r}$ that $\frac{d\phi }{dt}=\frac{2MJ}{r^{3}}$. This implies that the
particle starting with an initial radial fall begins to co-rotate with the
black hole in its vicinity (Lense-Thirring effect). In general, we shall
take $L\neq 0$. Conventionally, the sign of $J$ is taken as positive or
negative according as the source rotation is in the counterclockwise or
clockwise sense.

To examine stability, we first note that the autonomous system is of the
same type as in Eqs.(15), (16), only the coefficients are different. The
next step is to follow the same procedure as in Sec.3. Applying it for a
particle starting at $v_{0}=0$, we get the equilibrium points in the ($%
x^{\prime },y^{\prime }$) plane at $Q_{1}:\left( 0,0\right) $ and $%
Q_{2}:\left( -\frac{\beta }{c},0\right) $ where%
\begin{equation}
\beta =\pm \sqrt{\left( 1+\frac{8JM^{2}}{L^{3}}\right) ^{2}-\frac{12M^{2}}{%
L^{2}}}
\end{equation}%
which reduces to $\alpha $ when $J=0$. The interesting result is that the
reality of $\beta $ immediately imposes two restrictions, viz., that $J\neq -%
\frac{L^{3}}{8M^{2}}$ and that the quantity under the radical sign in
Eq.(83) must be positive which implies $(J-J_{+})(J-J_{-})>0\Rightarrow $
either $J<J_{\pm }$ or $J>J_{\pm }$. These restrictions must be respected if
circular orbits are to exist at all in Kerr spacetime. Once this is
fulfilled, exactly the same arguments about the stability as in Sec.3 go
through under the replacement of $\alpha $ by $\beta $.

\textit{Case (a)}: $\beta ^{2}=0$

This degenerate condition corresponds to two critical values $J_{+}$, $J_{-}$
of $J$ which are 
\begin{equation}
J_{\pm }=\mp \frac{L^{3}}{8M^{2}}\left[ 2\sqrt{3}\frac{M}{L}\pm 1\right] .
\end{equation}%
The phase paths are the same as those given by Eq.(25) indicating unstable
equilibrium at radii given by%
\begin{equation}
r_{\beta =0}^{Kerr}=\frac{6M}{1+\frac{8JM^{2}}{L^{3}}}.
\end{equation}%
Putting the values of $J_{\pm }$ from Eq.(84), we find the radii $r_{\beta
=0}^{Kerr}=\mp \sqrt{3}L$, \ which implies that the rotation $J$ of the
source has no role in determining the radii of circular orbits if $\beta
^{2}=0$.

\textit{Case (b)}: $\beta ^{2}>0$.

The phase paths are the same as Eq.(36) with the replacement of $\alpha $ by 
$\beta $. Hamiltonian analysis reveals that $Q_{1}$ is a center and $Q_{2}$
is a saddle if $\beta <0$. These conclusions are reversed if $\beta >0$.
Stable circular orbits occur at%
\begin{equation}
r_{\beta \neq 0}^{Kerr}=\frac{6M}{(1+\frac{8JM^{2}}{L^{3}})-\sqrt{\left( 1+%
\frac{8JM^{2}}{L^{3}}\right) ^{2}-\frac{12M^{2}}{L^{2}}}}.
\end{equation}%
From the above, it follows that, as $J\rightarrow \pm \infty $, the stable
radii $r_{\beta \neq 0}^{Kerr}$ go far beyond $6M$. This implies that the
rotation $J$ of the source can not bring about stable circular orbits at
radii below $6M$ for material test particles. The situation is the same as
in the nonrotating case.

\textit{Case (c)}: $a\neq 0$.

For light ($v_{0}=c_{0}$), $a=\frac{-2MJ}{L^{3}}\neq 0$, hence the
equilibrium points occur at $R_{1}:\left( 0,0\right) $ and $R_{2}:\left( -%
\frac{\sigma }{c},0\right) $ where 
\begin{equation}
\sigma =\pm \sqrt{1+\frac{24JM^{2}}{L^{3}}}.
\end{equation}%
Arguments similar to \textit{Case (a), }Sec.3 go through, with the
replacement of $\alpha $ by $\sigma $. Thus stable radii occur at%
\begin{equation}
r_{Light}^{Kerr}=\frac{6M}{1-\sqrt{1+\frac{24JM^{2}}{L^{3}}}}
\end{equation}%
depending on the sign of $\sigma $. When $J=0$, the orbit has infinite
radius. When $-1\leq \frac{24M^{2}J}{L^{3}}<0$, stable circular orbits will
exist for $r_{Light}^{Kerr}\geq 6M$ but when $\frac{24M^{2}J}{L^{3}}>0$,
there can not be any stable radius for light because $r_{Light}^{Kerr}$
becomes negative.

\begin{center}
\textbf{7. Optical black holes}
\end{center}

The advantage of the phase space method is that it can be applied to
situations beyond known gravitation theory when a potential function is not
always evident. This can occur, for instance, when one deals with a path
equation in the environment of a simulated black hole described by a
refractive medium with index $n(r)$. Possibility of laboratory creation of
such optical black holes exist in view of a remarkable experiment [9]
performed in Bose-Einstein condensates. The experiment demonstrated that
optical pulses can travel in the condensate with extremely small group
velocities, as low as $17m/s$. The group velocity of light in the vicinity
of a real gravitational black hole can indeed be arbitrarily low [10]. It
has been shown that light motion around a dielectric vortex structure mimics
motion around a black hole [11]. Creation of an event horizon would require
that the vortex flow be supplemented with a radial flow as well [12].
Interesting physical effects, like optical Aharonov-Bohm effect far away
from the vortex core and bending of light near the core, stem from the
consideration of a dielectric medium having a velocity field $%
\overrightarrow{u}$ and a varying index of refraction $n$.

One might obtain the trajectory of light directly from Fermat's principle%
\begin{equation}
\delta \int n(\overrightarrow{r})dl=0.
\end{equation}%
The resulting path equation is given by Eq.(79) which, on the equatorial
plane $\theta =\pi /2$, gives%
\begin{equation}
\frac{dr}{d\phi }=\pm \frac{\lbrack r^{4}n^{2}(r)-r_{i}^{2}r^{2}]^{\frac{1}{2%
}}}{r_{i}}
\end{equation}%
where $r_{i}$ is a constant of integration, interpreted as impact parameter.
Stability of circular orbits can be easily studied directly once a form for $%
n(r)$ is given.

A plausible form for $n(r)$ simulating a static dielectric medium of optical
black holes has been studied by Marklund, Anderson, Cattani, Lisak and
Lundgren [13]. It is given by%
\begin{equation}
n^{2}(r)=1+\frac{r_{0}^{2}}{r^{2}}
\end{equation}%
which has a divergence at $r=0$, and $r_{0}$ is a constant. For $r\gg r_{0}$%
, $n(r)\sim 1$ and for $r\ll r_{0}$, $n(r)\sim \frac{r_{0}}{r}$. Defining,
as before, $u=x=\frac{1}{r}$, the autonomous system corresponding to the
problem can be written as%
\begin{eqnarray}
\overset{.}{x} &=&y \\
\overset{.}{y} &=&\varepsilon x,\varepsilon =\frac{r_{0}^{2}-r_{i}^{2}}{%
r_{i}^{2}}.
\end{eqnarray}%
The equilibrium point ($0,0$) refers to circular orbit only at the
asymptotic region $r=\infty $. Taking the limit $r_{i}\rightarrow \infty $,
we find $\varepsilon =-1$. Hence the path equation $y^{2}-\varepsilon
x^{2}=C $ represents a family of concentric ellipses around the origin
showing that the orbit is stable independent of any finite value of the
extent $r_{0}$ of inhomogeneity. But \textit{no} circular orbit at a finite
radius is possible. This result is quite consistent with the nature of
various trajectories analyzed in Ref.[13]. However, a pathological solution
of $\overset{.}{y}=0$ may be imagined by taking $\varepsilon =0\Rightarrow
r_{0}^{2}=r_{i}^{2}$. We find that there is no equilibrium point at all in
this case as $\overset{.}{x}\neq 0$ although $\overset{.}{y}=0$. The latter
yields a phase path equation $y=\frac{dx}{d\phi }=\sqrt{C}$ which integrates
to give real space trajectory $x\varpropto \phi $. This is just in the form
of Archimedes' spiral $r_{i}/r=\phi $ as shown in Ref.[13].

We show now that circular orbits are possible in a nonuniformly \textit{%
moving }medium with a slowly varying refractive index. Under these
conditions, Leonhardt and Piwnicki [11] considered a vortex core with a
velocity profile decaying away from the core%
\begin{equation}
\overset{\rightarrow }{u}=\frac{W}{r}e_{\widehat{\phi }}
\end{equation}%
where $2\pi W$ is the vorticity. Let us formally introduce the index given
by Eq.(91) into the Hamilton-Jacobi equation for light motion derived in
[11]. The resulting path equation can be translated to the following
autonomous system in the far field limit%
\begin{eqnarray}
\overset{.}{x} &=&y \\
\overset{.}{y} &=&-x+\frac{1}{l_{AB}^{2}}\left(
r_{0}^{2}x+2r_{0}^{2}W^{2}x^{3}+3r_{0}^{4}W^{2}x^{5}\right)
\end{eqnarray}%
where $l_{AB}$ is the Aharonov-Bohm modified angular momentum given by%
\begin{equation}
l_{AB}=l+(n^{2}-1)W=l+\frac{r_{0}^{2}}{r^{2}}W.
\end{equation}%
Interestingly, we find that there is only one equilibrium point (apart from
the trivial one at $x=r^{-1}=0$ or $r=\infty $) at a finite radius photon
orbit 
\begin{equation}
R=\sqrt{3}r_{0}\left[ -1+\left\{ 1+\frac{3(l_{AB}^{2}-r_{0}^{2})}{W^{2}}%
\right\} ^{\frac{1}{2}}\right] ^{-\frac{1}{2}}.
\end{equation}%
In order that this radius be real, we must have%
\begin{equation}
1+\frac{3(l_{AB}^{2}-r_{0}^{2})}{W^{2}}\equiv N^{2}>1
\end{equation}%
where $N$ is real. The quantity $q_{0}$ in this case works out to%
\begin{equation}
q_{0}=-\left( \frac{4}{3}\right) \frac{W^{2}}{l_{AB}^{2}}N(1-N)>0
\end{equation}%
if $N>1$. Thus the orbit is stable. Of course, the conclusion crucially
depends on the form of $n(r)$.

We finally mention an interesting similarity between $L$ of Eq.(82) and $%
l_{AB}$ of Eq.(97). A light ray approaching radially ($l=0$) will acquire an
angular momentum $(n^{2}-1)W$ near the vortex core very similar to the
Lense-Thirring effect. With $n^{2}\sim 1+\frac{4m}{r}$, where $m$ is some
constant, we have $\frac{d\phi }{dt}\sim \frac{4mW}{r^{3}}$. This allows us
to identify $2W$ as the angular momentum of the vortex motion.

\begin{center}
\textbf{8. Summary}
\end{center}

Phase space analysis has been successfully applied to practically all walks
of life, from physics, engineering, biology to social sciences. Somehow its
use in gravitational physics seems rather scarce. Our motivation here was to
fill that gap. There is certainly room for further development like
exploring how other sophisticated techniques from the phase space repertoire
could be applied to stability of noncircular orbits or even classical fields.

We obtained information on stability of circular orbits arguing from the
geometrical path equation alone. The usefulness of the method is
demonstrated in several situations of physical interest. When the
cosmological constant $\lambda =0$, the dimensionless parameter $\alpha $
played a key role in determining the stability of actual orbits in the
physical ($r,\phi $) space. It was shown that $\alpha =0$ is a cusp
describing a marginal state, viz., the radius $R=6M$ is neither stable nor
unstable. The light orbit at $R=3M$ is unstable independent of the sign of $%
\alpha $ and in the asymptotic region it is always stable. All the
conclusions were confirmed by the method of Hamiltonian system.

When $\lambda \neq 0$, there occurred either one or three equilibrium
points. The one corresponding to static radius does not depend on the
parameter $\gamma $ but depends on $\lambda $. We found that, for $\lambda
>0 $, trajectories of circular material orbit at the static radius are
unstable. However, for light orbits, $h\rightarrow \infty $, so that $%
q_{0}=1 $, hence circular light orbits at static radius are \textit{always}
stable. These two results help us understand better the nature of the static
hypersurface:\ Even though forces balance at the hypersurface, it is not
exactly like the usual flat asymptotic region where both matter and light
orbits are stable. The other two equilibrium points are local and already
analyzed in Sec.4. They do not depend on $\lambda $ implying that circular
orbits in the vicinity of $M$ are not influenced by $\lambda $. This is a
physically consistent result. It was shown how a single parameter $q_{0}$
nicely reproduced all the relevant results about circular orbits in the
Schwarzschild-de Sitter spacetime.

We dealt with circular motion of light and massive particles in the
equatorial plane of the Kerr black hole. The derived results are new. The
restrictions $J<J_{\pm }$ or $J>J_{\pm }$ for the existence of circular
orbits were derived. A general parameter $\beta $ was found that reduces to $%
\alpha $ under zero rotation, $J=0$. For the value $\beta ^{2}=0$, the
rotation $J$ of the source has no role in determining the radii of circular
orbits. For $\beta ^{2}>0$, we found that even high values of source
rotation can not bring about a stable radius below $6M$. These results could
be of importance to accretion phenomenon in astrophysics.

Finally, we applied the method to real optical dielectric (static and
moving) and obtained consistent results. Here we only discussed a simple
example relevant to optical black holes but \textit{any }given form of
refractive index $n(r)$ can be similarly handled. The important advantage is
that we did not require information on dynamical potential functions, but
relied solely on the path equations coming from Fermat's principle or
Hamilton-Jacobi equation. Several such path equations corresponding to
various refractive media have been worked out in Ref.[4]. It will naturally
be of interest to apply the method in the refractive wormhole
\textquotedblleft media" constructed from exotic matter [14] or in
Brans-Dicke theory [15,16]. Work is underway.

\textbf{Figure captions}

Fig.1. The origin $O$ is a cusp [See p.19, Ref.1]. Path $AO$ leads to the
origin while $OB$ leads away from $O$. These paths correspond to $C=0$. For $%
C\neq 0$, the paths $CDE$ never reach $O$ and the motion is analogous to the
whirling motion of the bob of a pendulum.

Fig.2. The origin $O$ is a stable center. Initial conditions slightly
shifted from the center take the phase paths on closed elliptic orbits
around $O$. The corresponding motion in physical space is periodic analogous
to small oscillations of the bob about downward vertical.

Fig.3. The origin $O$ is a saddle point. Only the paths $AO$ and $BO$
approach the origin while $OC$ and $OD$ move away from it. Other paths do
not lead to the origin. $EF$, $GH$ represent whirling motion.

\textbf{Acknowlegments}

The authors wish to thank Denis V. Kondratiev and Guzel N. Kutdusova for
technical assistance. KKN acknowledges warm hospitality at JRL where part of
the work is carried out.

\textbf{References}

[1] Jordan D W and Smith P 1999 \textit{Nonlinear Ordinary Differential
Equations} 3rd edition (Oxford University Press: Oxford)

[2] Reiss A G \textit{et al} 1998 Observational evidence from Supernovae for
an accelerating universe and a cosmological constant \textit{Astron. J}. 
\textbf{116 }1009-38; Garnavich P M \textit{et al} 1998 Supernova limits on
the cosmic equation of state \textit{Astrophys.J.} \textbf{509} 74-9;
Perlmutter S J \textit{et al }1998 Discovery of a supernova explosion at
half the age of the universe and its cosmological implications \textit{%
Nature (London)} \textbf{391} 51-4

[3] Stuchl\'{\i}k Z and Hled\'{\i}k S 1999 Some properties of the
Schwarzschild-de Sitter and Schwarzschild-anti de Sitter spacetimes \textit{%
Phys Rev.} D \textbf{60} 044006-15

[4] Evans J and Rosenquist M 1986 \textquotedblleft $F=ma$" optics \textit{%
Amer. J. Phys. }\textbf{54} 876-83\ 

[5] Evans J, Nandi K K and Islam A 1996 \ The optical-mechanical analogy in
general relativity:\ Exact Newtonian forms for the equations of motion of
particles and photons \textit{Gen. Relat. Grav.} \textbf{28} 413-39

[6] Nandi K K and Islam A 1995 On the optical-mechanical analogy in general
relativity \textit{Amer. J. Phys.} \textbf{63} 251-6

[7] Nandi K K, Migranov N G, Evans J and Amedeker M K 2005 Planetary and
light motions from Newtonian theory:\ An amusing exercise \textit{Eur. J.
Phys. }\textbf{27} 429-35

[8] Alsing P M 1998 The optical-mechanical analogy for stationary metrics in
general relativity \textit{Amer. J. Phys.} \ \textbf{66} 779-90

[9] Hau L V, Harris S E, Dutton Z\ and Behroozi C\ H 1999 Light speed
reduction to 17 metres per second in an ultracold atomic gas \textit{Nature
(London)} \textbf{397} 594-8

[10] Nandi K K, Zhang Y Z, Alsing P\ M, Evans J and Bhadra A 2003 Analogue
of the Fizeau effect in an effective optical medium \textit{Phys. Rev.} D 
\textbf{67} 025002-11

[11] Leonhardt U and Piwnicki P 1999 Optics of nonuniformly moving media 
\textit{Phys. Rev.} A \textbf{60}, 4301-12 ; Leonhardt U and Piwnicki P 2000
Relativistic effects of light in moving media with extremely low group
velocity \textit{Phys. Rev. Lett.} \textbf{84} 822-5

[12] Visser M 2000 Comments on \textquotedblleft Relativistic effects of
light in moving media with extremely low group velocity" \textit{Phys. Rev.
Lett.} \textbf{85} 5252

[13] Marklund M, Anderson D, Cattani F, Lisak M and Lundgren L 2002 Fermat's
principle and variational analysis of an optical model for light propagation
exhibiting a critical radius \textit{Amer. J. Phys. }\textbf{70} 680-3

[14] Ellis H\ G 1973 Ether flow through a drainhole:\ A particle model in
general relativity \textit{J. Math. Phys.} \textbf{14}, 104-8; Ellis H\ G
1974 Errata \textit{J. Math. Phys.}\textbf{15} 520

[15] Nandi K K, Islam A and Evans J 1997 Brans wormholes \textit{Phys. Rev.}
D \textbf{55} 2497-500

[16] Nandi K\ K, Bhattacharjee B, Alam S\ M\ K and Evans J 1998 Brans-Dicke
wormholes in the Jordan and Einstein frames \textit{Phys. Rev. }D \textbf{57}
823-28

\end{document}